# Exploring Unpopular Presidential Elections

Michael G. Neubauer
Mark Schilling
Joel Zeitlin
California State University, Northridge

*"That the people of England, being at this day very unequally distributed by Counties, Cities, and Boroughs for the election of their deputies in Parliament, ought to be more indifferently proportioned according to the number of the inhabitants; the circumstances whereof for number, place, and manner are to be set down before the end of this present Parliament."*

John Lilburne, 1647

## Historical Frequency of "Unpopular" Elections

In the 2000 presidential election, George W. Bush was elected by winning the electoral vote although Al Gore received the most popular votes. While this possibility, which we shall call an "unpopularly elected president", had always been recognized, it suddenly passed from an event not seen in anyone's lifetime and which might have been dismissed as highly unlikely, to an actual modern day occurrence that brought criticism of our democratic institutions, especially the Electoral College. Many felt the disagreement between popular vote and electoral vote was not fair or democratic—it had violated the notion of "one man, one vote". The smaller number of popular votes that Bush received were said to have counted for more than Gore's total, which exceeded Bush's by roughly one half million votes--yet Bush won the Electoral College vote. (One of the electors pledged to Gore did not vote for Gore and so the actual result was 271-266.)

The dissatisfaction with the discrepancies between the Electoral College outcome and the popular vote that occur form time to time has spawned a movement in support of The National Popular Vote Bill [http://www.nationalpopularvote.com], which would effectively guarantee the Presidency to the winner of the national popular vote once enough states adopt the bill to guarantee victory for the national popular winner. In that case the adopting states then award all their electors to the national popular winner. As of May 2011 the bill had been adopted in nine states representing 49% of the electoral votes needed to activate it.

Gallup polls dating as far back as 1944 have consistently shown only about 20% of the public supporting the current electoral system and about 70% opposed. Recent surveys have found more than 70% favor a direct nationwide election of the President. While the public's feelings about the appropriateness of the Electoral College system may have been intensified by recent elections, instances where the winner of the Presidency did not win at least a plurality of the popular vote are in fact common in our nation's history:

> 1. In 1824, John Quincy Adams was elected with 32% of the popular vote and only 84 out of a total of 261 electoral votes while Andrew Jackson won 42% of the popular vote and 99 electoral votes. Two other candidates, William Crawford and Henry Clay, received 41 and 37 electoral votes respectively. The U.S. Constitution requires a majority, not just a plurality, of Electoral College votes for a candidate to become president. In case no candidate has a majority of the Electoral College votes the U.S.

Constitution prescribes that the House of Representatives decides among the top three electoral vote finishers; each state has only one vote in the matter. On the first ballot in the House Adams won 13 states to Jackson's seven and Clay's four.

2. In 1876 Rutherford B. Hayes (Republican) became President with 48% of the popular vote and 185 electoral votes while Samuel J. Tilden (Democrat) won 51% of the popular vote but only 184 electoral votes. Interestingly, the apportionment of the House of Representatives following the previous (1870) census was not carried out according to the accepted apportionment method of the time (known as *Hamilton's method*) or any other recognized method and represented a compromise reached in the House. If Hamilton's method had been used, as was the law at the time, Tilden would have won the Electoral College vote by one vote.

3. In 1888, Benjamin Harrison (Republican) became President with 48% of the popular vote and 233 electoral votes while incumbent Grover Cleveland (Democrat) received 49% of the popular vote but just 168 electoral votes.

4. Most recently George W. Bush was elected in 2000 with 271 electoral votes to Al Gore's 266 while receiving roughly one half million fewer popular votes than Gore.

5. A rather ambiguous example occurred in 1960 when John F. Kennedy defeated Richard M. Nixon by 84 Electoral votes. Due to a complicated situation in Mississippi and Alabama regarding unpledged delegates on the ballot, there is some question about how to properly count popular votes for Kennedy in these states. (See the article by Gaines in *Further Reading* for a discussion of this election and the difficulty in listing vote counts for Kennedy.) It has been argued that votes for these unpledged delegates should not be credited as popular votes for Kennedy, in which case Kennedy did not win a plurality of the popular vote.

Thus for the 58 presidential elections that have taken place since 1780, at least four unpopular elections—nearly 7%—have occurred. If we include the 1960 election, fully 8.6% of all American presidential elections have been unpopular.

Estimates based on the historical record do not necessarily provide good predictions of the likelihood of an unpopular result in a future election because the sample size involved in these estimates is small, and because the political structure has changed dramatically many times in United States history. A more effective way to estimate the likelihood of a future unpopular election is to use the *Monte Carlo* method: simulate a large number of elections using a suitable randomization approach and then compute the percentage that result in unpopularly elected presidents, using as input data an appropriate set of recent presidential elections.

### Structure of the Electoral College

The number of electors for each state is the number of representatives for that state, plus two more for the Senate seats. The House size was set at 435 in 1911 and has not been changed since then. The District of Columbia was granted three electoral votes in 1961 by the 23$^{rd}$ Amendment, thus there are currently $435 + 50 \times 2 + 3 = 538$ members of the Electoral College and this has been so for the last 12 presidential elections, starting in 1964. We regard District of Columbia as a state for the purposes of this article, which results in 102 "Senate" electors and 436 "House" electors.

**Simulating Presidential Elections**

For the purposes of this study we awarded fixed numbers of electors using the "winner take all" method in all states even though two states, Nebraska and Maine, do otherwise. Since together they have only nine electoral votes and usually all electoral votes go to the same candidate in both states, this simplifying assumption does not greatly influence the validity of our model. We based our simulations on the data for the last twelve presidential elections (1964 through 2008) to avail ourselves of the fullest history that represents current conditions.

One might wonder if, for example, the political dynamics of the Nixon versus Kennedy election are significantly different than more recent elections and so the elections from many years ago might not be germane to our analysis. With this in mind we ran several simulations that began in election years later than 1964. The results for these other analyses were similar, so we do not provide those results here. We excluded votes for third party candidates and regard each potential election as between a Republican and a Democrat. All of the data we use was taken from the Statistical Abstracts of the United States, published by the US Census Bureau.

One simple idea for simulating elections is to choose for each of the 51 "states" with equal likelihood, one of the past twelve elections. Then using that year's percentages of votes for Democrats and Republicans out of the total votes cast for those two parties, determine which party receives the resulting electoral votes. Next, multiply these percentages by current voting population figures for that state to obtain the popular votes for each party for that state. Summing this procedure over all states provides the electoral and popular vote totals for one simulated election.

This simulation method is unsatisfactory, however; while it will reproduce the state marginal distributions of the actual data, the joint distribution will be the product of these 51 marginal distributions. However the actual election data is not independent from one state to another, but in fact is highly correlated; for example, the southern states east of Texas often vote in much the same way and in so doing have had a strong voice in determining the president. Other, less geographically connected states also have highly correlated voting patterns, as seen in Figure 1. Over half of the state-to-state correlations for the 1964-2008 election records are above 0.70; this reflects the substantial dependency that exists among state voting patterns. To adequately simulate presidential elections, this dependency must be take into account.

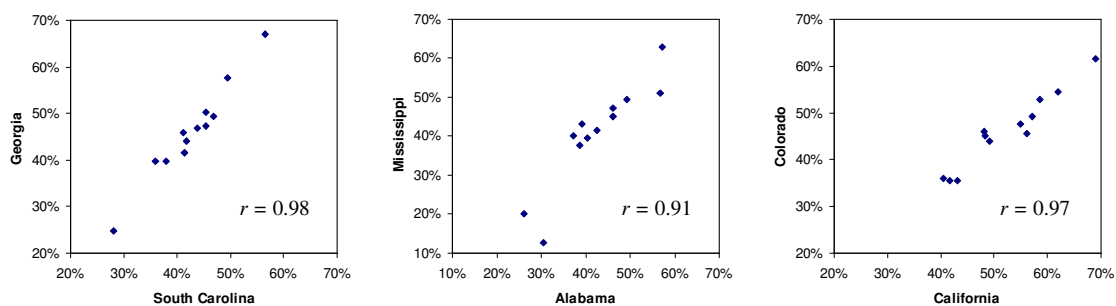

Figure 1. Many state voting patterns exhibit strong positive correlation.

We incorporated the historical correlation between states' voting patterns by means of *principal components analysis.* (See the Appendix for a description of principal components analysis and details of how we implemented it for this study.) We represented the data from the last twelve elections as twelve 51-dimensional vectors of measurements, where the 51 components consisted of each state's percentage of votes for the Democratic candidate out of the total of Republican and Democratic votes. We used the eigenvectors and eigenvalues generated by the principal component analysis to compute simulated percentages of Republican and Democratic votes for each state as follows:

Let $\lambda_j, j = 1, \ldots, 11$ represent the non-zero eigenvalues of the sample covariance matrix of the data and let $E_j, j = 1, \ldots, 11$ be the corresponding eigenvectors. Also let $z_j$, for $j = 1, \ldots, 11$ be independent normally distributed random variables with mean zero and variance one. Then a simulated vector of state election percentages is generated by the formula

$$\mu + \sum_{j=1}^{11} z_j \sqrt{\lambda_j} \ E_j \ .$$

This method generates multivariate normal 51-dimensional vectors with mean vector and variance-covariance matrix matching that of the actual data for the past twelve elections. Note that $\sqrt{\lambda_j}$ represents the standard deviation of the data in the direction of the eigenvector $c_j$; scaling by this factor thus gives each eigenvector the proper weight in the random linear combination above. Multiplying the simulated state percentages by the numbers of voters in each state in 2008 gives simulated popular vote totals, which then determine electoral votes awarded to each party. Each new generation of eleven independent standard normal random variables {$z_j, j = 1, \ldots, 11$} gives a different simulated election.

The values of the eleven principal components are listed in Table 1. Their sizes diminish rapidly, indicating that the data structure is dominated by the first few eigenvectors (especially the first two or three) and that the impact of the last several principal components is relatively unimportant.

| 1 | 2 | 3 | 4 | 5 | 6 | 7 | 8 | 9 | 10 | 11 |
|---|---|---|---|---|---|---|---|---|---|---|
| 0.2217 | 0.0512 | 0.0316 | 0.0109 | 0.0086 | 0.0063 | 0.0039 | 0.0033 | 0.0019 | 0.0006 | 0.0004 |

Table 1. Eigenvalues of the non-zero eigenvectors.

A side benefit of using principal components to simulate presidential elections is that insights into the structure of recent presidential election history are gained. In Table 2 the coefficients (or coordinates) for each state for each of the first three principal components have been sorted by size and listed next to their state name. These show the contribution of each state to the given factor. The eigenvector corresponding to the largest eigenvalue represents the direction of maximal variation of the set of twelve points in 51-dimensional space, and the corresponding eigenvalue in Table 1 is the amount of variation (variance) in that direction.

| eigenvec1 | state | eigenvec2 | state | eigenvec3 | state |
|---|---|---|---|---|---|
| -0.0238415 | Mississippi | -0.121403 | California | -0.260707 | Vermont |
| 0.0186228 | Alabama | -0.116453 | Alaska | -0.248836 | Jersey New |
| 0.0452776 | DC | -0.0919417 | Vermont | -0.2418 | Mississippi |
| 0.0701703 | Carolina South | -0.0881028 | Hampshire New | -0.185957 | DC |
| 0.0799757 | Louisiana | -0.0829933 | Oregon | -0.178365 | Illinois |
| 0.0871298 | Dakota South | -0.0794905 | Colorado | -0.166086 | Delaware |
| 0.0895441 | Minnesota | -0.0764787 | Massachusetts | -0.16361 | New York |
| 0.0902731 | Georgia | -0.0749953 | Utah | -0.142217 | Connecticut |
| 0.100288 | Wisconsin | -0.0740215 | Connecticut | -0.135817 | Florida |
| 0.110931 | Tennessee | -0.0713542 | Dakota North | -0.129319 | Maryland |
| 0.111254 | Iowa | -0.0688837 | Nevada | -0.114116 | Hampshire New |
| 0.119318 | Oregon | -0.0663339 | Dakota South | -0.112617 | Arizona |
| 0.119699 | Montana | -0.0660958 | Iowa | -0.0940777 | Nevada |
| 0.119978 | Kansas | -0.0633663 | Nebraska | -0.0928782 | Hawaii |
| 0.122596 | Indiana | -0.0579008 | Michigan | -0.0863222 | California |
| 0.126893 | Carolina North | -0.0531243 | Montana | -0.0782796 | Washington |
| 0.127318 | Virginia | -0.0523939 | Hawaii | -0.0766953 | Mexico New |
| 0.128131 | Arizona | -0.0484016 | Wisconsin | -0.0757859 | Virginia |
| 0.129364 | Kentucky | -0.0472812 | Washington | -0.0730579 | Maine |
| 0.129483 | Pennsylvania | -0.0389907 | Mexico New | -0.0416486 | Michigan |
| 0.130103 | Idaho | -0.0357221 | Idaho | -0.0338265 | Massachusetts |
| 0.130325 | Nebraska | -0.033154 | New York | -0.0301716 | Carolina South |
| 0.132685 | Illinois | -0.0300273 | Island Rhode | -0.0263555 | Island Rhode |
| 0.132848 | Dakota North | -0.0284695 | Illinois | -0.0138041 | Colorado |
| 0.133248 | Ohio | -0.0247437 | Arizona | -0.00946684 | Alabama |
| 0.136501 | Washington | -0.012697 | Wyoming | -0.00908988 | Louisiana |
| 0.139055 | Virginia West | -0.00979605 | Pennsylvania | -0.00258206 | Oregon |
| 0.141333 | Missouri | -0.008557 | Ohio | 0.00597608 | Wisconsin |
| 0.142502 | Arkansas | 0.000857517 | Maine | 0.00932561 | Iowa |
| 0.142899 | Mexico New | 0.00100859 | Jersey New | 0.00988972 | Pennsylvania |
| 0.143523 | Texas | 0.00202122 | DC | 0.0219484 | Ohio |
| 0.14354 | Wyoming | 0.00742675 | Indiana | 0.0324945 | Indiana |
| 0.144142 | Alaska | 0.0128174 | Delaware | 0.0367796 | Georgia |
| 0.144723 | Maryland | 0.017701 | Maryland | 0.0440678 | Carolina North |
| 0.145804 | Florida | 0.0188452 | Kansas | 0.073195 | Tennessee |
| 0.146466 | Delaware | 0.0310493 | Minnesota | 0.0743097 | Minnesota |
| 0.149133 | Michigan | 0.0336506 | Missouri | 0.0877877 | Missouri |
| 0.149201 | Oklahoma | 0.0637789 | Virginia | 0.097551 | Nebraska |
| 0.151662 | Massachusetts | 0.0657446 | Texas | 0.124997 | Arkansas |
| 0.154299 | Utah | 0.0887935 | Kentucky | 0.132984 | Kansas |
| 0.158538 | Colorado | 0.113207 | Oklahoma | 0.160389 | Montana |
| 0.165692 | New York | 0.123198 | Florida | 0.160393 | Idaho |
| 0.169652 | California | 0.136418 | Virginia West | 0.166403 | Dakota North |
| 0.172565 | Nevada | 0.139861 | Carolina North | 0.171023 | Dakota South |
| 0.173609 | Connecticut | 0.175465 | Tennessee | 0.185169 | Oklahoma |
| 0.18331 | Island Rhode | 0.257683 | Carolina South | 0.20485 | Utah |
| 0.184728 | Maine | 0.261201 | Louisiana | 0.222316 | Texas |
| 0.190381 | Vermont | 0.279381 | Arkansas | 0.222965 | Kentucky |
| 0.202345 | Hampshire New | 0.380418 | Georgia | 0.241793 | Virginia West |
| 0.220533 | Hawaii | 0.396645 | Alabama | 0.267344 | Wyoming |
| 0.222753 | Jersey New | 0.501711 | Mississippi | 0.277031 | Alaska |

Table 2. Coefficients of the first three eigenvectors sorted by size.

Note that for all but one state, Mississippi, the coefficients of the first principal component, eigenvec1, have the same sign. This reflects the fact that almost all of the states have tended to vote together overall. This makes sense since a candidate who wins in a landslide will tend to receive a higher proportion of votes in nearly every state than will a candidate in a closer election. The highest positive coefficients of the second principal component, eigenvec2, are all southern states, confirming notions that their behavior is correlated and tends to be different from that of the remaining states. The second most significant factor in the past twelve elections is therefore the southern/non-southern political axis.

Finally, the third principal component, eigenvec3, appears to separate urban and rural states to some degree. These first three factors (overall popularity, southern vs. non-southern support, urban vs. rural support) explain fully 88% of the variation in the data (computed from the sum of the $1^{st}$ three eigenvalues divided by the sum of all of them).

**Estimated Frequency of Unpopular Elections**

The primary goal of our simulation study was to estimate the frequency of unpopular elections under reasonably current conditions. We used the principal components approach to generate 20,000 trials and found a 4.9% frequency of unpopular elections. Although the estimated likelihood of an unpopular election suggests that such an outcome would occur on average only once every eighty years, the fact that there have already been at least four such outcomes in our nation's history suggests that the issue is nevertheless an important one.

The Democratic candidate won 44% of the simulated elections and the Republican candidate won 56%, which closely matches the actual proportions for the 1964-2008 elections, of which Democrats won five and Republicans seven. This provides confirming evidence that our simulation approach reflects presidential voting patterns for the period 1964-2008.

**Effect of Awarding Senate Electors to Each State**

When the founding fathers designed our electoral voting system, they decided to award each state two electors (representing their two senators) in addition to the electors they received according to their representation in the House of Representatives. The examples in Table 3 below show how "unpopular" outcomes can occur both (i) without and (ii) with the extra Senate electors in the Electoral College. We describe each case by a two place code in which in the first position indicates whether the popular winner wins (W) or loses (L) the sum of the House and Senate electors (our present electoral system), and the second position indicates whether the popular winner wins (W) or loses (L) the majority of House electors only.

For simplicity let's pretend there are only three states having populations of 300, 100 and 100 and they are apportioned to have three, one and one House seats respectively. After adding the Senate electors they then have five, three and three electors. In all scenarios below, A wins the popular vote.

**LW, unpopular election "caused" by Senate electors**

| State | A's % | Popular Vote | | | House + Senate Electoral Votes | | | House Electoral Votes | |
|---|---|---|---|---|---|---|---|---|---|
| | | A | B | | A | B | | A | B |
| 1 | 53% | 159 | 141 | | 5 | | | 3 | |
| 2 | 47% | 47 | 53 | | | 3 | | | 1 |
| 3 | 47% | 47 | 53 | | | 3 | | | 1 |
| Total | | **253** | 247 | | 5 | **6** | | **3** | 2 |

**LL, unpopular election using either method**

| State | A's % | Popular Vote | | | House + Senate Electoral Votes | | | House Electoral Votes | |
|---|---|---|---|---|---|---|---|---|---|
| | | A | B | | A | B | | A | B |
| 1 | 49% | 147 | 153 | | | 5 | | | 3 |
| 2 | 49% | 49 | 51 | | | 3 | | | 1 |
| 3 | 65% | 65 | 35 | | 3 | | | 1 | |
| Total | | **261** | 239 | | 3 | **8** | | 1 | **4** |

**WL, Unpopular election only when using only House electors**

| State | A's % | Popular Vote | | | House + Senate Electoral Votes | | | House Electoral Votes | |
|---|---|---|---|---|---|---|---|---|---|
| | | A | B | | A | B | | A | B |
| 1 | 49% | 147 | 153 | | | 5 | | | 3 |
| 2 | 53% | 53 | 47 | | 3 | | | 1 | |
| 3 | 53% | 53 | 47 | | 3 | | | 1 | |
| Total | | **253** | 247 | | **6** | 5 | | 2 | **3** |

Table 3. Simplified scenarios yielding different unpopular results.

In the first scenario, LW (like the Bush versus Gore election), candidate A wins the popular vote and loses the electoral vote with both House and Senate electors and wins the electoral vote without Senate electors This election turned out unpopular with both House and Senate electors counted because: (i) the margin of victory/defeat is the same in all three states and A wins the large state, which has a larger population than the other two states combined, so A wins the popular vote; and (ii) B triumphs in the Electoral College because B is rewarded for winning more states.

In the second scenario, LL, A wins the popular vote and loses both the full vote of the Electoral College (with Senate electors) and the House electoral vote alone. The reason this occurs is that A's margin of defeat in States 1 and 2 is slight, therefore A still wins the popular vote; but losing two states gives the majority of electoral votes to B, with or without the Senate electors.

In the third scenario, WL, A closely loses State 1 but wins more substantially in the other two states, so that A wins the popular vote but loses when the two states' Senate electors are not counted.

Clearly we can have disagreement between popular winner and electoral winner whether we include the Senate electors in the Electoral College or not. If a candidate wins many states by a narrow popular vote margin while losing others by large margins, it is certainly possible for that candidate to be "unpopularly elected" with a minority of the popular vote. Thus unpopular presidents will inevitably occur with a "winner take all" system for the states (as the U.S. currently has with the exception of Maine and Nebraska).

The number of electors due to House representation is approximately proportional to state populations, but with the addition of the Senate electors this proportionality relationship is weakened. This naturally leads to the question, "Is the frequency of unpopular elections higher with the added Senate electors or without?"

A priori, an argument can be made for either outcome: On the one hand, since including the Senate electors results in the number of electors apportioned to each state being less proportional to state populations, we might expect that Electoral College totals would less accurately reflect the overall popular vote than if only the electors due to the House were used; this would then result in more unpopular elections with Senate electors than would occur without them.

A counterargument is that the total number $S$ of Senate electors a candidate wins is positively correlated with the percentage of the popular vote he or she receives, just as is the total number $H$ of electors the candidate receives due to their House electors. Having two factors that are each correlated with the popular vote strengthens the association between the total electoral vote and the popular vote. Therefore our present system, which uses both $H$ and $S$ (giving $H + S$ electors to the candidate), will produce results better aligned with the popular vote than if $H$ alone were used.

Our simulations indicate a clear answer to this question. Figure 2 plots the number of state electors $S$ won by the candidate who won the popular vote (which is $2 \times$ the number of states won) against the number of House electors $H$ that candidate won. We show the results for only 2500 simulated elections in order to keep the density of the points shown manageable.

Remembering that considering the District of Columbia as a state with two Senate electors gives a Senate of size 102 and House of size 436, we see that points falling to the left of the vertical line $H = 436/2 = 218$ represent simulated outcomes that would be unpopular if the electoral college were based only on the House electors. Points falling to the lower left of the slanted line $H + S = 538/2 = 269$ correspond to simulated elections that would be unpopular under the present electoral system. The most obvious aspect of Figure 2 is that most elections are *not* unpopular. As indicated above, roughly 5% of our simulated elections are unpopular under the present system. Figure 2 also shows, not surprisingly, that the great majority of unpopular elections would be unpopular under either system (based either on $H$ or $H + S$).

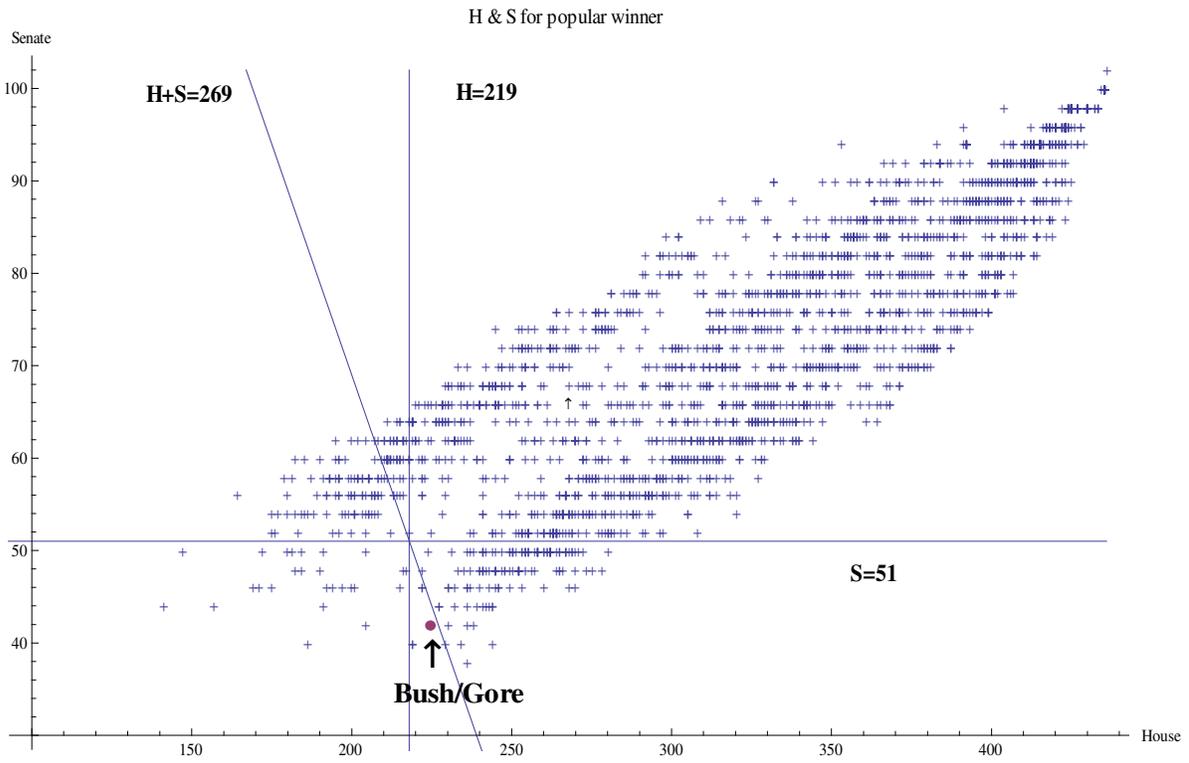

Figure 2. Scatter plot of 2500 elections showing total House and Senate electoral votes for the popular winner. The Bush/Gore outcome, (225,42), is also shown.

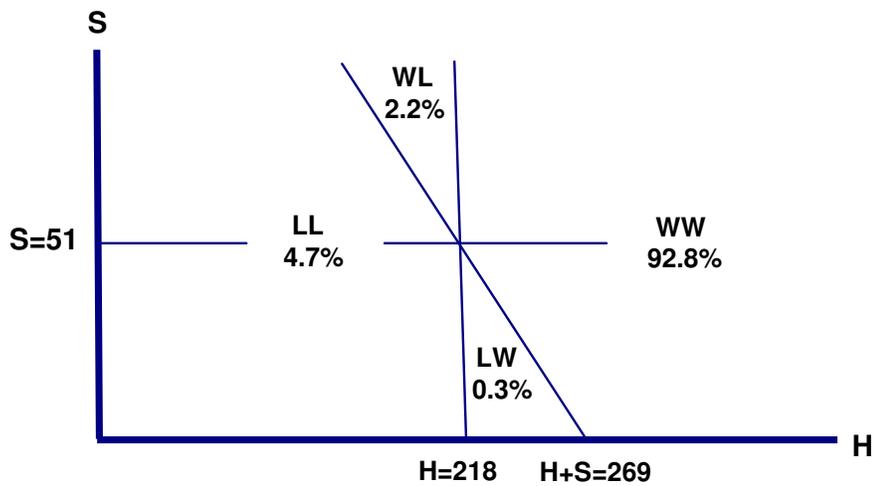

Figure 3. Regions corresponding to WL, LL and LW.

In order to decide which system produces fewer unpopular elections, it suffices to compare the incidence of points in the two wedge shaped regions corresponding to WL+LL (popular winner loses in the House) and LW+LL (popular winner loses with the Senate included), or more simply we can just compare WL and LW. As indicated in the figure, far more simulations fall in the WL region than the LW region, which means that unpopular elections are more common when just the House electors are used than when both the House and Senate electors are counted. Thus granting the extra two votes for Senate electors significantly lessens the likelihood that the electoral vote will differ from the popular vote. In our simulations, the popular vote differed from the full electoral vote in 0.3 % + 4.7 % = 5.0 % of the trials (LW+LL) and the popular vote differed from the House vote in 2.2 % + 4.7 %=  6.9 % of the trials (WL+LL).

One can also simulate elections under alternative rules that add larger numbers of Senate electors to the number of House-based electoral votes for each state. Note that increasing the number of Senate electors for each state to some large number would essentially result in awarding the presidency to the candidate winning the greater number of states.

We can get a good idea of how increasing the number of Senate electors would affect the frequency of unpopular elections by looking again at the slanted line in Figure 2. Giving more than two Senate electors for each state is tantamount to flattening the line; as the number of Senate electors increases, the line rotates counterclockwise towards its limiting position, the horizontal line in Figure 2. (The vertical axis would be scaled with different values, but that is immaterial.) It's clear that the frequency of unpopular elections (indicated by the proportion of points to the lower left of the slanted line) does not change dramatically as the line rotates; even when the number of Senate electors is so large that the winner is the candidate that wins more than half of the states, the estimated likelihood of an unpopular election is 6.1%, only about 1% higher than what we found under the current system, and lower than with House electors alone. This is apparent from Figure 2 in that the number of points below the line $S = 51$ appears just slightly larger than the number of points to the lower left of the line $H + S = 269$.

### The National Popular Vote Bill

For another variation recall the National Popular Vote bill mentioned earlier, which would go into effect only when enough states have signed on so that they command a majority of electoral votes and thus guarantee that the national popular vote winner is the electoral winner. What if states instead took action before they commanded a majority?

It is easy to imagine that unpopular elections tend to be fairly close. Figure 4, however, shows the frequency of the differences of electoral votes (Democratic electoral votes – Republican electoral votes) for all of the unpopular elections under the current system that occurred in our 20,000 simulations.

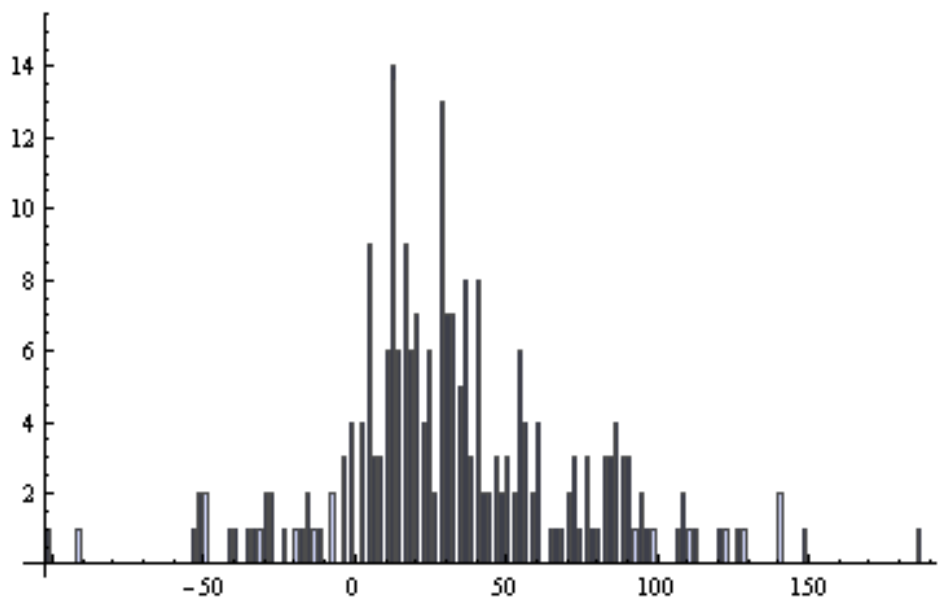

Figure 4. Frequency distribution of signed electoral differences,

Democrats electoral votes – Republican electoral votes.

Clearly electoral differences in unpopular elections can often be quite large and so the strategy of awarding states' electoral votes to the popular winner would not reliably guarantee the popular winner the election unless states representing a very large number of electoral votes are committed to it. Also note in Figure 4 that the majority of the electoral differences are positive, which indicates that if future voting patterns are like those of the past 50 years, the Democratic candidate is more likely to be the beneficiary of an unpopular election than the Republican.

The primary reason for this is that Republican candidates tend to have sizeable majorities in states they win, although most are small states. This boosts their overall popular vote without necessarily giving them enough large states to win the electoral vote. The most striking illustration of the consequences of failing to win in large states is what might be called the *California effect*. Figure 5 codes each simulated election according to whether the popular winner was (i) a Democrat or a Republican, and (ii) whether they carried California. Note that nearly all of the points in the cluster that lies in the upper left quadrant are cases in which the Republican was the popular winner but failed to capture California (and lost the electoral vote). As indicated in figure 3 above, the points below the slanted line H+S = 269 represent unpopular outcomes.

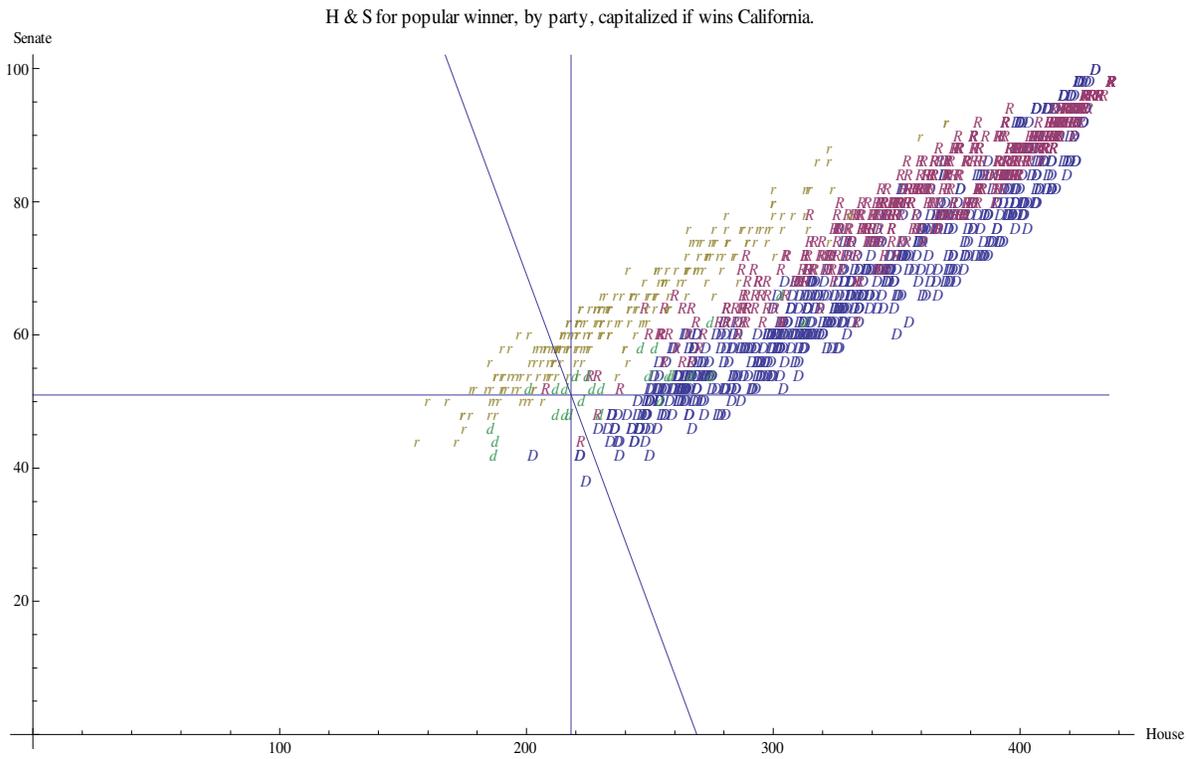

Figure 5. Scatter plot of 1500 elections showing total House and Senate electoral votes for the popular winner, by party of popular winner and capitalized if the popular winner wins California.

## Final Thoughts: The Evolution of the Electoral College

Lastly, we comment that the Electoral College at its inception was not supposed to reflect popular will. Alexander Hamilton in Number 68 of the Federalist Papers made it quite clear that a popularly elected President was not the intention of the framers. Such an important decision as electing the president, in Hamilton's words, was to be placed in the hands of "*men most capable of analyzing the qualities adapted to the station, and acting under circumstances favorable to deliberation, and to a judicious combination of all the reasons and inducements which were proper to govern their choice. A small number of persons, selected by their fellow-citizens from the general mass, will be most likely to possess the information and discernment requisite to such complicated investigations.*" Hamilton might find the point of departure for this paper very much inappropriate and contrary to the very idea of the Electoral College. Over time, however, presidential elections have become more like popular elections as political parties evolved and more states transferred the right "*to appoint* (electors), *in such manner as the Legislature thereof may direct*" to the people of the states by allowing them to vote for the electors directly. Today, in many states the names of electors are not even printed on the ballot anymore; only the candidates for President and Vice President are listed on the ballots. The Electoral College today is functioning very differently than originally imagined by the framers of the constitution.

**Further Reading**

R. Erickson, K. Sigman: "A Simple Stochastic Model for Close U.S. Presidential Elections", Department of Political Science, Columbia University, New York. (http://www.columbia.edu/~ks20/Erik-sig-academic.pdf)

COMAP, *For All Practical Purposes*, 8[th] ed. W.H. Freeman, New York, 2009. See Chapter 12, *Electing the President.*

B. Gaines, Popular Myths about Popular Vote-Electoral College Splits, *PS: Political Science and Politics*, **34 (**1) 70-71 (Mar 2001).

R. Johnston, D. Rossiter, C. Pattie, Disproportionality and Bias in the Results of the 2005 General Election in Great Britain: Evaluating the Electoral System's Impact, *Journal of Elections, Public Opinion & Parties*, 16 (1) 37-54 (2006).

*E. Kaplan and A. Barnett,* A New Approach to Estimating the Probability of Winning the Presidency, *Operations Research*, **51** (1), 32-40 (Jan-Feb 2003).

M. Neubauer and J. Zeitlin, Outcomes of Presidential Elections and the House Size, *PS: Political Science and Politics*, **36** 721 –725 (Oct 2003).

**Appendix: Principal Components Analysis Details**

We wish to exploit the correlations that exist between states' voting patterns to determine the core structure of the elections data. We represent all the data from the last twelve elections as twelve 51-dimensional vectors of measurements, where the 51 components represent each state's percentages for the Democratic candidate out of the total of Republican and Democratic votes. Subtracting the mean of the twelve percentages to center the data produces a $12 \times 51$ matrix of deviations which we call Devs. The rank of Devs is at most 11, not 12, since the deviations for each state sum to zero. We write $\Sigma = \text{Devs}^T \text{Devs}$, a $51 \times 51$ matrix of rank at most 11 or $n$-1. $\Sigma/(n\text{-}1)$ is the sample covariance matrix of the state variables; that is, the $(i,j)$-th element of $\Sigma/(n\text{-}1)$ is the sample covariance of the deviations of the $i$th and $j$th states.

The **principal components** are the eleven unit eigenvectors of $\Sigma/(n\text{-}1)$. The eigenvector corresponding to the largest eigenvalue represents the direction of maximal variation of the set of twelve points in 51-dimensional space, and the eigenvalue is the amount of such variation. More precisely, the largest eigenvalue is the variance of the data when it is projected onto the first eigenvector. The eigenvector corresponding to the second largest eigenvalue is the direction of maximal variation of the set of points in the space orthogonal to the first eigenvector (one can think of reducing the dimension of the point cloud by "projecting out" the first direction ), etc. Each eigenvector is a linear combination of the 51-dimensional basis vectors (1,0,0,…,0), (0,1,0,…,0),…, (0,...,0,1) associated with the 51 states, with coefficients (known as "factor loadings") reflecting the strength of the contribution of each state. These eleven eigenvectors are orthonormal.

Our principal components analysis could have been done using an eigenanalysis of the correlation matrix rather than the covariance matrix $\Sigma$. We chose the covariance matrix for this application since the 51 variables are commensurable, each measuring the same quantity for a different state.